\documentclass[preprint2]{aastex}

\newcommand{\uas}{$\mu${as}}
\slugcomment{Accepted December 4, 2008 to PASP}

\shorttitle{SIM Narrow Angle Performance}
\shortauthors{Shao et al.}

\begin{document}

\title{Sub-Microarcsecond Astrometry with SIM-Lite: \\
    A Testbed-based Performance Assessment }

\author{M. Shao\altaffilmark{1}, B. Nemati\altaffilmark{1} }
\affil{JPL, California Institute of Technology, MS 301-486, 4800 Oak Grove Drive, Pasadena, CA 91109}

\begin{abstract}

SIM-Lite is an astrometric interferometer being designed for sub-microarcsecond astrometry, with a wide range of applications from searches for Earth-analogs to determining the distribution of dark matter. SIM-Lite measurements can be limited by random and systematic errors, as well as astrophysical noise. In this paper we focus on instrument systematic errors and report  results from SIM-Lite's interferometer testbed. We find that, for narrow-angle astrometry such as used for planet finding, the end-of-mission noise floor for SIM-Lite is below 0.035\ \uas.

\end{abstract}


\section{Introduction}

The newly redesigned light version of the Space Interferometry Mission, SIM-Lite, is a long-baseline astrometric interferometer designed to reach sub-microarcsecond precision over the course of a five year mission. It has two modes of operation, one for global astrometry with 4~\uas\ end-of-mission accuracy, and the other for narrow-angle astrometry, with a single-epoch (1100s visit) accuracy below 1~\uas\ \citep{Unwin08}. The redesign lowers costs primarily by shortening the baseline from $9~{\rm m}$ to $6~{\rm m}$ and replacing a guide interferometer with a telescope star tracker.

Narrow-angle astrometry is used to measure the orbital motions of non-luminous objects, like neutron stars and planets, by observing the motion of the parent star.  SIM-Lite is designed to make narrow angle measurements by alternately switching between a target star and anywhere from 3 to 6 reference stars within a 1 degree radius. The astrometric signal is the motion of the target star relative to the reference stars.

To take a concrete example, consider a candidate star that is observed 250 times over the course of a 5-year mission.  A peak in the joint periodogram power distribution (similar to a power spectral density) of the target star position constitutes the detection of a planet candidate. Detection of a planet with a 1\% false-alarm probability requires a signal-to-noise ratio (SNR) of approximately 5.8. At 10 pc, the astrometric signature for an Earth orbiting a one Solar mass star at 1 AU is 0.3~\uas, so the SNR requirement calls for a final instrument error of less than 50~nano-arcseconds (nas). This corresponds to a single-epoch error of less than 0.82~\uas. For the nearest $\sim60$ stars, a noise floor below 35 nas is needed to detect Earths in the habitable zone.

The key questions, therefore, are whether sub-microarcsecond single-epoch accuracy is attainable, and, equally important, whether the instrument systematic errors do average down well below the single-epoch accuracy. This paper reports testbed results relating directly to both the single-epoch and end-of-mission narrow angle accuracy.

\begin{figure}
\epsscale{0.90}
\plotone{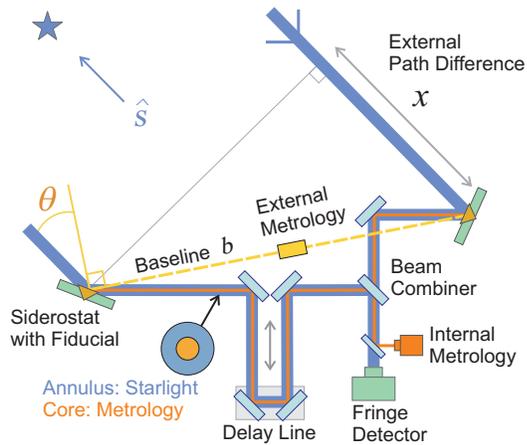}
\caption{The basic elements of a Michelson stellar interferometer.
\label{fig1}}
\end{figure}

The basic elements of a Michelson stellar interferometer are shown in Figure~\ref{fig1}. Starlight is collected by two spatially separated siderostats. Each siderostat has an embedded fiducial that marks one end of the baseline and also acts as a retroreflector for the internal and external metrology laser beams. Along each arm, the starlight beam has an annular footprint while internal metrology occupies the center.

The quantity of interest is the delay $x$, which is the optical path difference (OPD) between two starlight beams originating at a star and terminating at the two fiducials. The delay is given by:
\begin{equation}
x=\vec{b}\cdot \hat{s} + C + \eta
\label{eq-basicAstrom}
\end{equation}
where $\vec{b}$ is the baseline vector, $\hat{s}$ is the unit vector to the star, and $\eta$ represents measurement noise. $C$ is sometimes called the interferometer {\em constant term} and represents the internal OPD when the metrology reading is zero. The baseline length $b$ is defined as the distance between the two fiducials and is monitored by external metrology. Thus, three measurements, the white light fringe delay, internal metrology and external metrology, form the basic ingredients of the astrometric angle.

The SIM-Lite instrument is described in detail elsewhere \citep{SIM08, SIMEng08}. The fundamental SIM-Lite instrument is its ``science" interferometer, operating in the visible range with $50~{\rm cm}$ collectors and a $6~{\rm m}$ baseline. The baseline vector is not stable at the microarcsec level, so guide interferometers observing bright (typically 7mag) guide stars are used to measure the motion of the baseline over time. A laser optical truss ties everything together at the microarcsec level. A detailed mathematical analysis of the SIM-Lite astrometric approach appears in  \citet*{Mil03}.

\section{Random versus Systematic Errors}

The noise in astrometric equation~(\ref{eq-basicAstrom}) consists of both random and systematic instrumental errors. For a 6m baseline and a star close to the center of the field of regard, Equation~(\ref{eq-basicAstrom})
implies that 30 pico-meters (pm) of total delay error will cause approximately 1~\uas\ of astrometric error.

Random errors, also called white noise, will average down with integration time $T$ as $1/\sqrt{T}$. Examples of white noise include photon noise from the target and reference stars and detector noise in the laser metrology. Systematic error, or pink noise, is correlated over time and may not get smaller with longer integration time. The dominant systematic errors in SIM-Lite are thermal in nature. Thermal drift in the motion of the optics is properly monitored to first order by the laser metrology system and doesn't produce an error. If the metrology for some reason doesn't measure the optical path of the starlight properly, there will be an error. The main cause of such an error is the drift of the laser metrology beam's alignment with respect to the starlight.

SIM-Lite actively aligns both the metrology beam and the starlight beam using separate tip/tilt sensors in the astrometric beam combiner. Misalignment is caused by temperature gradients in the beam combiner that cause the metrology tip/tilt sensor to move relative to the starlight tip/tilt sensor. Only changes in the hardware that cause the metrology to incorrectly measure the optical path of the starlight are errors. Because the metrology hits the center of all the optics, while the star light has an annular footprint, warping of an optic can cause an error.

The best way to ensure we have captured all the important errors is a hardware testbed that has all the essential elements of the instrument we plan to operate in space. The Micro-Arcsecond Metrology (MAM) testbed was built for this purpose. Among other things, it captures the beam pathlength and angle control and beam recombination in a traceable manner to SIM-Lite.

\section{Results from the Interferometer Testbed}

The MAM testbed \citep{Hin02} has two main components: a test article corresponding to SIM-Lite, and an inverse interferometer pseudo star (IIPS)
to simulate the incoming star light. Starting from the IIPS source, broadband (600-1000 nm) light is injected from a fiber tip, collimated, and separated into two beams. These are steered to two coordinated stages where they are launched towards the test article siderostats as flat, coherent wavefronts. The test article contains all the essential components of the actual interferometer, including fringe detection,
pathlength and pointing control, and internal metrology \citep{An05}. The testbed was in operation from 2002 through 2006 and has been of
great value in identifying technical challenges, their mitigation, and demonstrating technical readiness \citep{GoulB02, GoShen04, Gou04}. 

\begin{figure}
\epsscale{.99}
\plotone{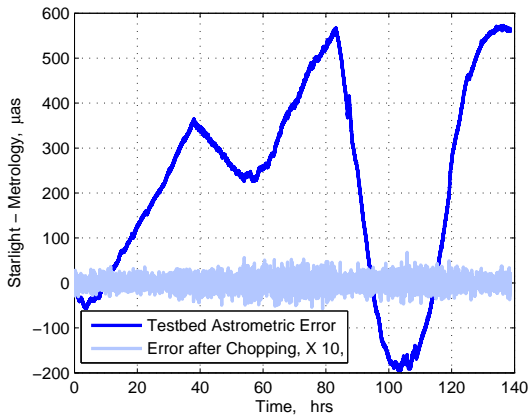}
\caption{140 hours of testbed data on the metrology-starlight error. Also shown (scaled up by x10) is the differential (chopped) error.
\label{fig2}}
\end{figure}

Figure~\ref{fig2} shows the effect of the metrology-starlight drift as measured in the testbed over a 140 hour period. To minimize the impact of thermal drifts on narrow-angle (NA) observations, SIM-Lite ``chops" between the target and each of the reference stars. The chop sequence is target, reference, target, next reference, target, and so on. While the actual integration times depend on stellar brightnesses, a ``typical'' chop between two stars consists of 15 seconds on the target, 30 seconds on the reference, and 15 seconds on slews between the stars. This ``differential" target-reference measurement error is also plotted in Figure~\ref{fig2}.

\begin{figure}
\epsscale{0.90}
\plotone{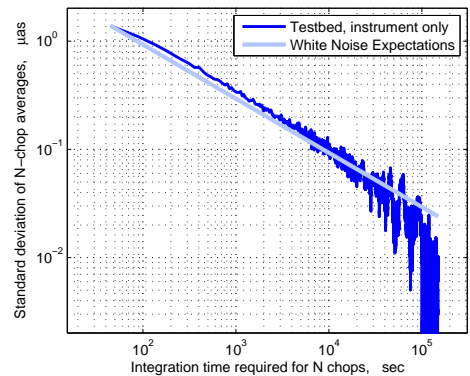}
\caption{The standard deviation $N$-chop averages as $N$ is increased, vs. integration time.  Testbed data are compared with white noise expectations.
\label{fig3}}
\end{figure}

Figure~\ref{fig3} shows the standard deviation of averages of $N$ chops versus the required integration time as $N$ is increased. We see that the standard deviation of the differential measurement decreases as $\sqrt{T}$, and from time scales of 45 seconds to 42 hours, the noise in the differential measurement is nearly white. No noise floor is observed down to 8 nas after about 42 hours. Since the last value is obviously a downward fluctuation, we extrapolate from the minimum integration time of 45 seconds to a white noise expectation of 24 nas at 42 hours.

An important question in applying testbed results to SIM-Lite is how does the thermal behavior of the relevant parts of SIM-Lite on orbit compare with our ground testbed. Within the testbed there were numerous temperature sensors that recorded the temperature fluctuations of the optics and mounts inside the vacuum chamber.

We also conducted a very detailed thermal simulation of the SIM-Lite spacecraft in orbit. This multi-thousand node thermal model was run for 100 hours of SIM-Lite operation, where the spacecraft attitude was allowed to change according a typical observing scenario. The simulation includes active thermal control, with the temperature maintained near $20^\circ$C within the expected capabilities of SIM-Lite's thermal control system. As different parts of the spacecraft are illuminated by the Sun, the simulated thermal control system turns heaters on and off in order to minimize temperature fluctuations in the key opto-mechanical environments.

\begin{figure}
\epsscale{0.95}
\plotone{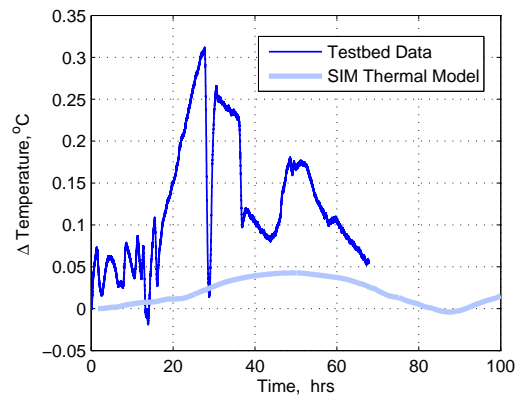}
\caption{Comparison of SIM thermal model with testbed thermal measurements.
\label{fig4}}
\end{figure}

Figure~\ref{fig4} shows the resulting predicted thermal fluctuations of the SIM-Lite beam combiner. Also shown in the figure are measured testbed environment temperatures under conditions similar to those that yielded the results of Figure~\ref{fig3}. We see that the actual thermal environment of SIM-Lite should be better than the testbed, or conversely, that the test results are conservative with respect to thermal drift.

\begin{figure}
\epsscale{0.95}
\plotone{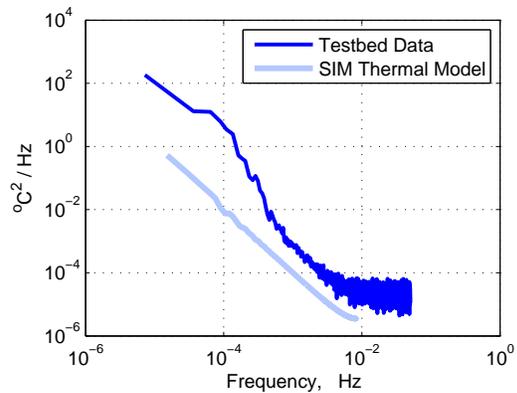}
\caption{Comparison of SIM thermal model with testbed thermal power spectral densities.
\label{fig5}}
\end{figure}

To directly compare the stability of SIM-Lite thermal environment to the testbed, we need to extract statistical quantities that describe the stability of the two thermal environments. One such quantity is the power spectrum of the thermal fluctuations, shown in Figure~\ref{fig5}.  SIM-Lite on orbit is seen to be more thermally stable than our current testbed environment.

\section{Astrophysical Sources of Error}

While this paper deals primarily with instrumental error sources, we have also extensively studied astrophysical errors which we summarize here. The two major astrophysical errors in astrometry are stellar activity and companions to reference stars.

A typical sunspot 0.1\% the area of the Sun will in the worst case shift the photocenter of the Sun by 0.25~\uas\ at 10 pc and produce a radial velocity bias of 1 ${\rm m}/{\rm s}$.
A shift of 0.25~\uas\ is well below the single-epoch precision of SIM-Lite. Star spot noise will not average down with integration time until that time is comparable to 1/4 - 1/2 the rotation period of the star or the mean sunspot lifetime.

The presence of planets orbiting reference stars is another astrophysical noise source. RV vetting of reference stars will eliminate binary stars and planets down to about Jupiter mass.  With multiple reference stars, it is possible to detect and assign the remaining planets to a specific reference star. A more detailed treatment will appear in a subsequent paper.

\section{Conclusion}

For planet finding, the dominant error experienced by an astrometric interferometer is the thermally induced drift of the metrology with respect to the starlight. The testbed results show that, even in the presence of a harsher thermal environment than SIM-Lite will experience in space, chopping between the target and reference stars renders this error nearly ``white'' with a noise floor below 24 nas after 42 hours of averaging. This is below the 35 nas required to detect an Earth in the habitable zone of the nearest $\sim60$ stars. While investigation of the next-level noise sources is continuing, these results suggest that the technology for astrometric detection of nearby Earths is at hand.

\section{Acknowledgements}

The research described in this paper was carried out at the Jet Propulsion Laboratory, California Institute of Technology, under a contract with the National Aeronautics and Space Administration.

Copyright 2008 California Institute of Technology. Government sponsorship acknowledged.

\clearpage

\end{document}